\documentclass{article}
\usepackage[nonatbib, final]{nips_2017}
\usepackage[utf8]{inputenc} 
\usepackage[T1]{fontenc}    
\usepackage{hyperref}       
\usepackage{url}            
\usepackage{booktabs}       
\usepackage{amsfonts}       
\usepackage{nicefrac}       
\usepackage{microtype}      
\usepackage{graphicx}
\usepackage{amsmath}
\usepackage{bm}

\title{Structural Vibration Signal Denoising \\ Using Stacking Ensemble of Hybrid CNN-RNN}

\author{Youzhi Liang \\
       Department of Computer Science\\
       Stanford University\\
       Stanford, CA 94305, USA \\
       \texttt{youzhil@stanford.edu} \\
       \AND
       Wen Liang \\
       Google Inc. \\
       Mountain View, CA 94043 \\
       \texttt{liangwen@google.com} \\
       \AND
       Jianguo Jia\\
       Department of Computing \\
       Hong Kong Polytechnic University\\
       Hong Kong, China \\
       \texttt{jianguo1.jia@connect.polyu.hk} }

\begin{document}

\maketitle

\begin{abstract}
Vibration signals have been increasingly utilized in various engineering fields for analysis and monitoring purposes, including structural health monitoring, fault diagnosis and damage detection, where vibration signals can provide valuable information about the condition and integrity of structures. In recent years, there has been a growing trend towards the use of vibration signals in the field of bioengineering. Activity-induced structural vibrations, particularly footstep-induced signals, are useful for analyzing the movement of biological systems such as the human body and animals, providing valuable information regarding an individual's gait, body mass, and posture, making them an attractive tool for health monitoring, security, and human-computer interaction. However, the presence of various types of noise can compromise the accuracy of footstep-induced signal analysis. In this paper, we propose a novel ensemble model that leverages both the ensemble of multiple signals and of recurrent and convolutional neural network predictions. The proposed model consists of three stages: preprocessing, hybrid modeling, and ensemble. In the preprocessing stage, features are extracted using the Fast Fourier Transform and wavelet transform to capture the underlying physics-governed dynamics of the system and extract spatial and temporal features. In the hybrid modeling stage, a bi-directional LSTM is used to denoise the noisy signal concatenated with FFT results, and a CNN is used to obtain a condensed feature representation of the signal. In the ensemble stage, three layers of a fully-connected neural network are used to produce the final denoised signal. The proposed model addresses the challenges associated with structural vibration signals, which outperforms the prevailing algorithms for a wide range of noise levels, evaluated using PSNR, SNR, and WMAPE.
\end{abstract}
\section{Background and Introduction}

The application of structural vibration signal is prevalent in various engineering domains, including civil, mechanical, and bioengineering. Its utility in civil engineering lies in monitoring the dynamic behavior of structures such as bridges and buildings subjected to loads like earthquakes and winds~\cite{speckmann2004structural}. Structural Health Monitoring (SHM) systems have been installed on such structures to ensure their safety and reliability, as aging structures may become a safety hazard~\cite{cross2013long, kaya2015real}. The SHM systems acquire structural responses to detect any abnormal conditions, allowing for timely maintenance and improved decision-making. Therefore, the accurate extraction of the underlying structural vibration signal from civil architectures is essential to provide a robust foundation for the reliability of SHM systems. In mechanical engineering, the use of structural vibration signal is vital in monitoring the health of machines, such as turbines, pumps, and engines~\cite{malekloo2022machine, fish2019dynamic, hoseinzadeh2018quantitative}. The effective monitoring of equipment condition and fault diagnosis is crucial to ensure safe operation. The accurate assessment of equipment condition necessitates high-quality data acquisition that contains significant structural vibration information and low noise levels~\cite{lei2020applications, wang2022attention}.

The use of vibration signals in bioengineering has seen a recent surge in popularity. Structural vibrations caused by activity provide a powerful tool for analyzing the movement of biological systems such as animals and humans~\cite{alyasseri2019eeg, galica2009subsensory, kessler2019vibration, hahm2022home}. The study of footstep-induced signals has emerged as a crucial area of research, given that these signals contain unique information on an individual's gait, body mass, and posture, which could used as a biometric feature to identify individuals~\cite{pan2015indoor, shoji2004personal}. Footstep-induced signals have broad-ranging applications in fields such as health monitoring, security, and human-computer interaction~\cite{ekimov2006vibration, li2019smart}. The analysis of these signals can provide useful insights into an individual's physical condition, such as early indicators of movement disorders or gait patterns indicating injuries or illnesses~\cite{fagert2019gait}. Additionally, these signals can be used to develop secure and non-intrusive biometric authentication systems, which are essential in modern security applications. Finally, the study of footstep-induced signals can also revolutionize the field of human-computer interaction, as it enables the development of new natural and intuitive interfaces between humans and machines~\cite{drira2021using}. Overall, the benefits of studying footstep-induced signals are enormous, as they have potential applications in a wide range of fields, including healthcare, security, and technology~\cite{fagert2020structural}. Denoising of these vibration signals is crucial for accurate analysis and interpretation of the underlying physical phenomena. Vibration signals are often corrupted by noise from various sources, such as electrical interference, environmental factors, and measurement errors. To address this issue, various signal processing techniques have been developed, such as Fourier transform, wavelet transform, and empirical mode decomposition~\cite{kumar2021stationary, kaur2021eeg}. However, these methods have limitations in handling complex and non-stationary signals with high noise levels~\cite{bayer2019iterative}. As such, there is a pressing need for denoising methods that are both robust and efficient in capturing the underlying structural mechanics that govern dynamic signals.

In this paper, we propose a novel architecture of a hybrid CNN-RNN stacking ensemble model, where the bi-directional LSTM architecture enables the model to capture both forward and backward temporal dependencies in the signal and the CNN architecture extracts the condensed representation of multiple vibration signals for the mutual denoising governed by complex dynamics. To model the underlying structural dynamics, we utilize a set of PDE/ODEs and evaluate our model by comparing its performance to that of other prevailing models using the synthetic dataset.

\section{Dataset}

\begin{figure}[ht]
  \includegraphics[width=125mm]{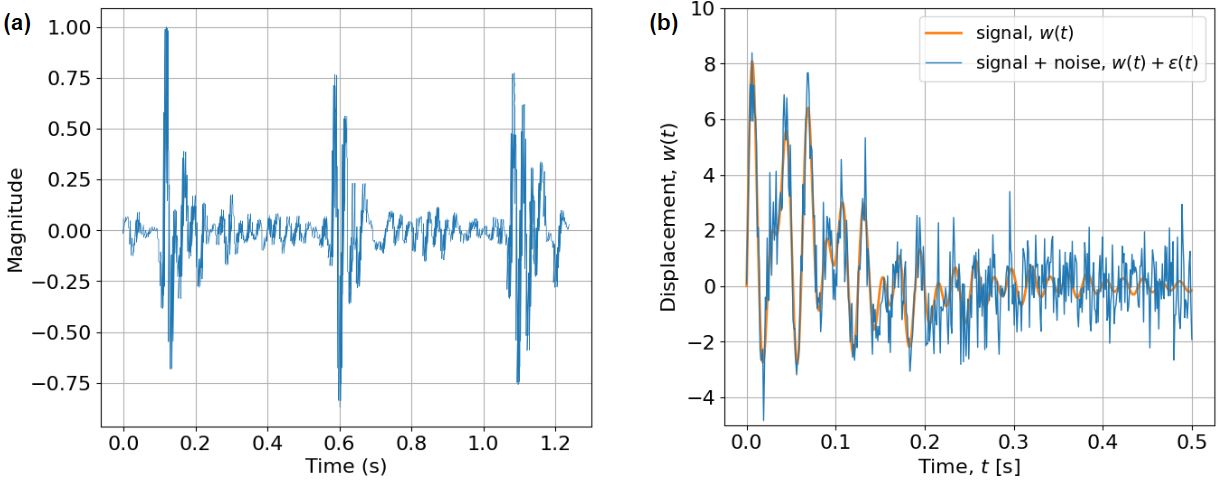}
  \centering
\caption{(a) A sample of foot-step induced floor vibration signal, normalized magnitude of vibration as a function of time~\cite{pan2017footprintid, hu2021footstep}. (b) An example of the signals generated based on Eqn.~\ref{eqn:ODE}, displacement overlaid with a high level of supplemental Gaussian noise, $w(t)$ as a function of time, $t$ [s].}
  \label{fig:sample_signal}
\end{figure}

The ground truth of a structural vibration signal can be challenging to obtain in practice due to various factors, such as the presence of noise, measurement errors, and limitations in the digitalization process~\cite{wang2022attention}. It is often impossible to measure or observe the true vibration signal directly, and hence, to generate a dataset of structural vibration time series, we opted to use a PDE/ODE solver (\href{https://docs.scipy.org/doc/scipy/reference/integrate.html#module-scipy.integrate}{scipy.integrate}) instead of conducting expensive experiments or computational-intensive simulations using Abacus. Our choice was motivated by the fact that simulations and experiments can be prohibitively expensive and time-consuming, whereas the use of a solver allowed us to quickly and efficiently generate a large dataset of 100,000 synthetic time series ($\mathcal{R}^{100,000 \times 500})$. These synthetic time series were generated by adding supplementary noise to the output of the PDE/ODE solver. For example, an added Gaussian noise can be modeled as $\epsilon(t) = \mathcal{N}(0, \sigma^2_{\epsilon})$, where the standard deviation $\sigma_{\epsilon}$ also represents the level of noise in this paper.

Our study focuses on the dynamics of structural vibrations induced by footsteps. Footsteps generate a transient load that causes the structure to vibrate. This vibration can be modeled as the response of a Kirchhoff-Love plate, which is a widely used model in structural dynamics~\cite{reddy2020integrated, asakura2014finite}. The Kirchhoff-Love plate theory assumes that the plate is thin compared to its length and width, and that it is subject to small deformations. Additionally, we assume the linearity of isotropic plates, meaning that the material properties of the plate are uniform in all directions. Under these assumptions, the induced vibrations can be modeled as the solutions of a system of PDEs. Nguyen {\em et.~al}~\cite{nguyen2021stable} provides a detailed description of the system of PDEs that we use to model the induced vibrations, where the system consists of two coupled partial differential equations that describe the displacement and rotation of the plate~\cite{asakura2014finite}. The boundary conditions are given by the Dirichlet condition, which specifies the displacement of the plate along its boundaries~\cite{iosifescu2009nonlinear}. By solving this system of PDEs, we are able to generate a realistic simulation of the vibrations induced by footsteps on a Kirchhoff-Love plate.

The dynamics of structural vibrations induced by footsteps can be simplified as the impulse response of a Kirchhoff-Love plate subject to Dirichlet boundary conditions~\cite{feldmann2009design}. Under the assumption of linearity for isotropic plates, we model the induced vibrations as the solutions of a system of PDEs~\cite{nguyen2021stable, mora2020virtual}:

\begin{equation}
    \label{eqn:ODE}
    D_i \nabla ^2 \nabla^2 \bm{w}_i(\bm{x}, t) - T_i \nabla ^2 \bm{w}_i(\bm{x}, t) = \delta(\bm{x}, t) - \rho_i h_i \bm{\ddot{w}}_i(\bm{x}, t) - K_i \bm{\dot{w}}_i(\bm{x}, t),
\end{equation}
where $\bm{w}_i(\bm{x}, t)$ is the transverse deflection, $D_i \sim \mathcal{N}(\mu_D, \sigma_D)$, $T_i \sim \mathcal{N}(\mu_T, \sigma_T)$, $\rho_i h_i \sim \mathcal{N}(\mu_{\rho h}, \sigma_{\rho h})$ and $K_i \sim \mathcal{U}(a_U, b_U)$ are all parameters of the structure, governing the dynamical response subject to impulse $\delta(\bm{x}, t)$. 

The transverse deflection $\bm{w}_i(\bm{x}, t)$, a function of measured coordinates in 2-dimensional spatial domain and in temporal domain, represents the underlying noise-free signal that the denoising model attempts to retrieve, where subscript $i$ corresponds to a simulating scenario with a set of isotropic plate parameters subjected to an external stimulus, under the assumption that the underlying noise-free signal is linearly proportional to the transverse deflection. Each synthetic time series of the noisy signal, $\tilde{\bm{w}}$ is obtained by $\sum_i \left( w_i(\bm{x}, t) + \sum_j \epsilon_{i, j}(\bm{x}, t) \right) $, where $\epsilon_{i, j}(\bm{x}, t)$ represents the supplementary noise, potentially originating from various noise sources, including electromagnetic interference, acoustic interference, and sensor digitization and calibration. To validate our synthetic data, we conducted additional comparisons between our synthetic data and measurements reported in prior studies~\cite{pan2017footprintid, hu2021footstep}. The sample of measured foot-step induced floor vibration signal is obtained by a data collection system, consisting of distributed vibration sensing nodes to capture the floor vibration, representative of structural vibrations and are precise in nature~\cite{pan2017footprintid, hu2021footstep}. The signals can be obtained from one or more locations; for each location, the sythetic PDEs can be reduced to ODEs. Figure~\ref{fig:sample_signal} (a) illustrates the floor vibration signal induced by three footsteps, indicating a superposition of vibrations from mediums with varied natural frequencies and damping ratios. Figure~\ref{fig:sample_signal} (b) shows a sample synthetic result which aligns closely with the actual measurements of foot-step induced floor vibrations.

\section{Methods, Results and Discussion}

\begin{figure}[t]
  \includegraphics[width=120mm]{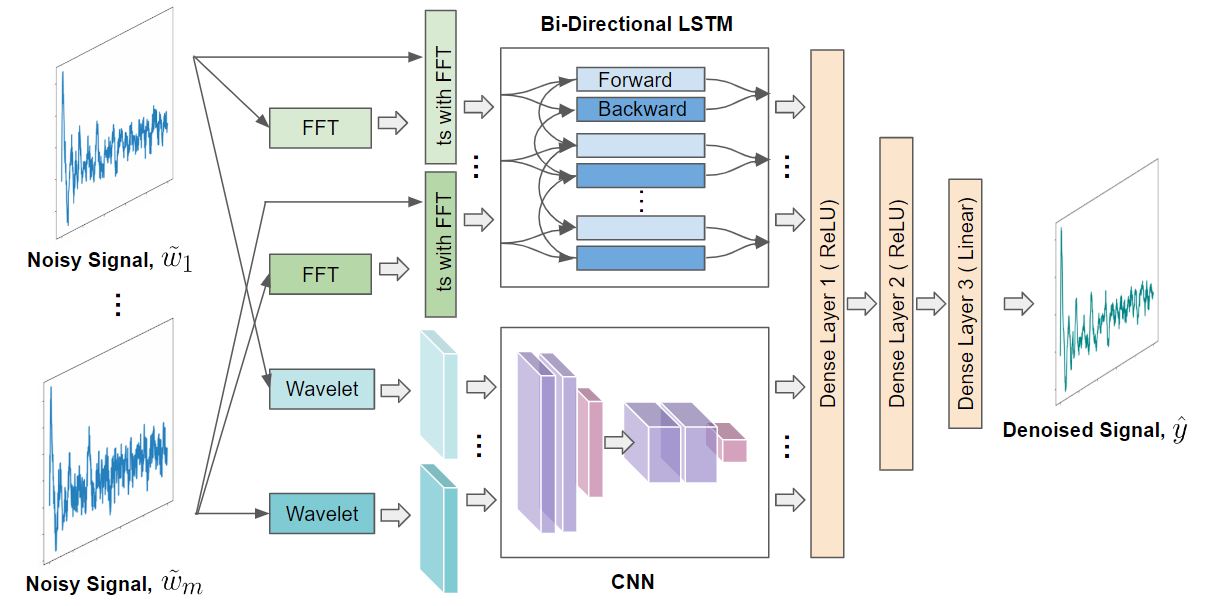}
  \centering
  \caption{A schematic of the architecture for our proposed stacking ensemble of hybrid CNN-RNN model. The input to the model can consist of multiple noisy signals, ranging from $\bm{\tilde{w}_1}$ to $\bm{\tilde{w}_m}$, while the output is a denoised signal, represented by $\bm{\hat{y}}$.}
  \label{fig:ensemble_architecture}
\end{figure}

We propose a stacking ensemble model that leverages both the ensemble of multiple signals and the predictions of Recurrent Neural Networks (RNNs) and Convolutional Neural Networks (CNNs). Our proposed model consists of three stages: a preprocessing stage, a hybrid modeling stage, and an ensemble stage, as depicted in Fig.~\ref{fig:ensemble_architecture}. Using multiple sensors is a common technique to improve the measurement accuracy of induced structural vibration signals. However, our proposed model is also able to handle the case where only a single measurement is available, allowing for effective denoising in this scenario as well. In the preprocessing stage, each signal is concatenated with its Fast Fourier Transform (FFT) results, while each image undergoes a set of wavelet transforms to extract spatial and temporal features. In the hybrid modeling stage, a bi-directional Long Short-Term Memory (LSTM) network is used to denoise the noisy signal concatenated with FFT results, and a CNN is used to extract a condensed feature representation of the signal. Finally, in the ensemble stage, three layers of fully-connected neural networks are used to produce the final denoised signal.

In the preprocessing stage, we employed multiple feature engineering techniques to extract useful features from the input data. Specifically, we used the Fast Fourier Transform (FFT) to extract features in the frequency domain, which capture the underlying physics-governed dynamics of the system and can help to improve the accuracy of predictions~\cite{lin2023fft, li2023research}. By incorporating these features as additional inputs into the bi-directional LSTM neural network, we were able to further improve the accuracy of our predictions. In addition to FFT-based features, we also applied wavelet transforms to extract spatial and temporal features from each signal in the input data. We used wavelets in the family of Daubechies and Bior to produce three channels of features in the wavelet transform domain~\cite{dautov2018wavelet}. Wavelet-based features are particularly useful for denoising in cases where nonstationary noise is present, enabling the decomposition of the noise from the signal~\cite{akansu2010emerging}. By combining FFT and wavelet-based features in our preprocessing stage, we were able to effectively capture both the spatial and temporal characteristics of the input data, leading to improved denoising performance in our model. Given that all the noisy signals originate from the same vibration source, a joint denoising utilizes the complementary and mutual underlying signal to reject the corrupted signal in presence of different levels of noise~\cite{wu2020multi}. The output of this first stage can be represented as follows:

\begin{align*}
    \bm{y}_{\textit{LSTM}}^{[1]} &= \bigcup_{i=1}^m \{ \bm{\tilde{w}}_i, \mathcal{F}_{\textit{FFT}} \left( \bm{\tilde{w}}_i \right) \}, \\
    \bm{y}_{\textit{CNN}}^{[1]} &= \bigcup_{i=1}^m \{ \mathcal{F}_{\textit{WT}} \left( \bm{\tilde{w}}_i \right) \},
\end{align*}
where $\bm{y}_{\textit{LSTM}}^{[1]}$ and  $\bm{y}_{\textit{CNN}}^{[1]}$ denote the outputs of the first layer serving as inputs for the following LSTM model and CNN model, respectively, $\mathcal{F}_{\textit{FFT}}$ and $\mathcal{F}_{\textit{CNN}}$ denote the Fast Fourier transform and Wavelet transform, respectively, and $i$ denotes the index of signal ranging from $0$ to $m$.

In the hybrid modeling stage, we employed a 'many-to-many' bi-directional LSTM architecture with a dropout rate of 0.2 to address the challenges associated with analyzing structural vibration signals, which are typically complex, non-stationary, and contaminated with noise (as shown in Fig.~\ref{fig:ensemble_architecture}). By using a bi-directional architecture, our model is able to capture both past and future temporal dependencies, which can improve the accuracy of predictions. Additionally, the use of dropout regularization helps to prevent overfitting and improve the generalization performance of the model. Simultaneously, each signal undergoes a set of wavelet transforms to extract spatial and temporal features, leading to improved performance in denoising structural vibration signals.. These features are stacked and used as input for a CNN, which consists of a sequence of convolutional layers with the same padding, followed by a ReLU activation layer and a max pooling layer. The resulting highly condensed features are then used to condition the ensemble in the next stage of our model~\cite{bai2019mental}. The output of this second stage can be expressed as:

\begin{align*}
    \bm{y}_{\textit{LSTM}}^{[2]} &= \mathcal{F}_{\textit{LSTM}} \left( \bm{y}_{\textit{LSTM}}^{[1]}; \bm{\Theta}_{\textit{LSTM}} \right), \\
    \bm{y}_{\textit{CNN}}^{[2]} &= \mathcal{F}_{\textit{CNN}} \left( \bm{y}_{\textit{CNN}}^{[1]}; \bm{\Theta}_{\textit{CNN}} \right),
\end{align*}
where $\mathcal{F}_{\textit{LSTM}} \left( \cdot; \bm{\Theta}_{LSTM} \right) $ and $\mathcal{F}_{\textit{CNN}} \left( \cdot; \bm{\Theta}_{CNN} \right) $ denote the functions of the bi-directional LSTM and CNN model, with trainable parameters $\bm{\Theta}_{LSTM}$ and $\bm{\Theta}_{CNN}$, respectively. 

In the ensemble stage, our proposed model consists of three fully connected neural network layers, each of which is followed by a rectified linear unit (ReLU) activation function, with a linear activation function applied to the final layer. The input to this stage is a concatenation of the embeddings from the bi-directional LSTM and CNN models, which have been applied to all noisy signals in the range of $\bm{\tilde{w}_1}$ to $\bm{\tilde{w}_m}$. The number of neurons in the last layer of the neural network is equal to the number of samples in each signal, with its output as the denoised signal, represented by $\bm{\hat{y}}$. The concatenation of the output results from the bi-directional LSTM and CNN models for all noisy signals can help to improve the overall accuracy of the ensemble predictions. In addition, we use wavelet features to condition the ensemble process. This approach helps to ensure that the model is able to effectively denoise each signal while still taking into account the unique characteristics of each signal. The output of the last stage can be obtained by 

$$\bm{\hat{y}} = \bm{y}^{[3]} = \mathcal{F}_{\textit{NN}} \left( \bm{y}_{\textit{LSTM}}^{[2]}, \bm{y}_{\textit{CNN}}^{[2]}; \bm{\Theta}_{\textit{NN}} \right),$$
where $\mathcal{F}_{\textit{NN}} \left( \cdot; \bm{\Theta}_{NN} \right) $ denotes the function of the fully-connected neural network, with trainable parameters $\bm{\Theta}_{NN}$. 

\begin{figure}[t]
  \includegraphics[width=130mm]{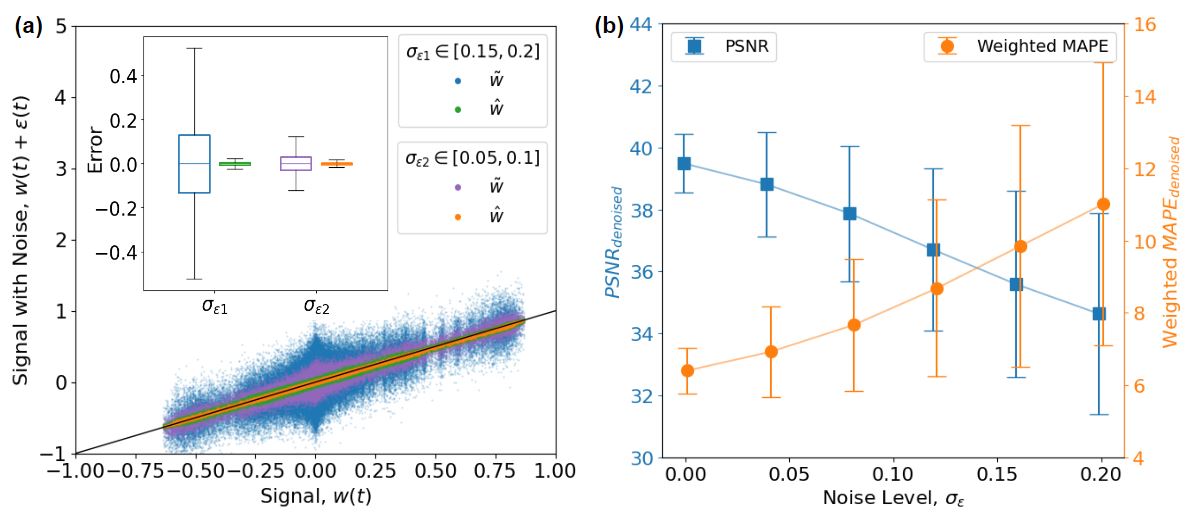}
  \centering
  \caption{(a) Noisy signals and denoised signals versus noise-free signals for level of noise $\sigma_{\epsilon} \in [0.05, 0.1]$ and $\sigma_{\epsilon} \in [0.15, 0.2]$. (Inset) Boxplots of error for noisy signals $\bm{\tilde{w}} - \bm{w}$ and for denoised signals $\bm{\hat{w}} - \bm{w}$ (b) PSNR of denoised signals and weighted MAPE of denoised signals versus noise level.}
  \label{fig:ensemble_result}
\end{figure}

We propose a novel cost function that aims at achieving a balance between model accuracy and generalization. To improve the accuracy of our model, we incorporate the expected loss over the training set using the $L_2$-norm. Furthermore, given that all the noisy signals originate from the same vibration source, we also include the difference between LSTM model predictions as a loss term. To enhance the generalizability of our model, we introduce regularization for all the trainable parameters of the LSTM, CNN, and NN models. The loss can be expressed as:

\begin{align*}
    \mathcal{L} \left( \bm{\Theta} \right) = &\mathbb{E}_{\bm{I}} \left\| \hat{\bm{y}} - {\bm{I}} \right\|^2 + \lambda_{\textit{LSTM}} \mathbb{E} \left\| \bm{\Theta}_{\textit{LSTM}} \right\|  + \lambda_{\textit{CNN}} \mathbb{E} \left\| \bm{\Theta}_{\textit{CNN}} \right\|  + \lambda_{\textit{NN}} \mathbb{E} \left\| \bm{\Theta}_{\textit{NN}} \right\|  \\ 
&+ \sum_{i \neq j} \lambda \mathbb{E} \left\| \bm{y}_{\textit{LSTM}}^{[2] \langle i \rangle } - \bm{y}_{\textit{LSTM}}^{[2] \langle j \rangle } \right\| ,
\end{align*}
where $\bm{y}_{\textit{LSTM}}^{[2] \langle i \rangle }$ represents the output of the second stage using LSTM model for the signal $i$.

\begin{figure}[t]
  \includegraphics[width=130mm]{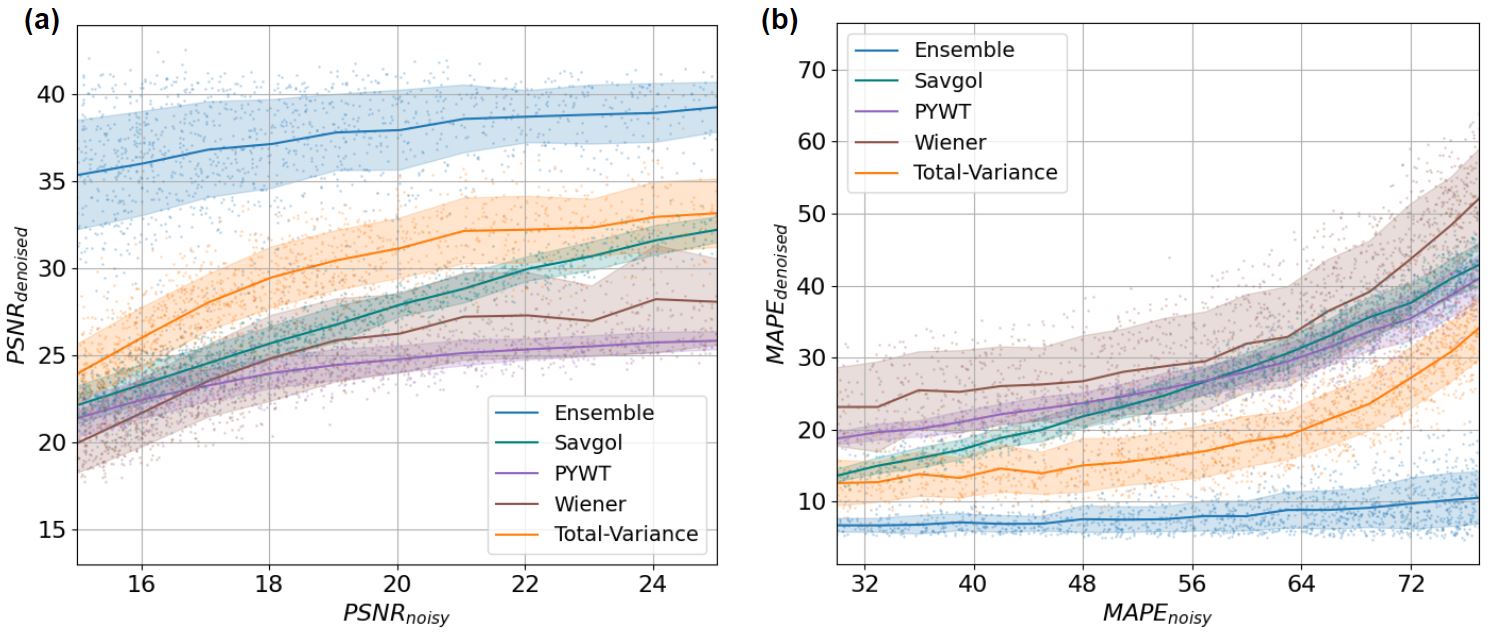}
  \centering
  \caption{(a) PSNR of denoised signals using five methods versus PSNR of noisy signals, where Ensemble denotes our model, and Savgol, PYWT, Wiener and Total-Variation denote Savitzky-Golay filter~\cite{press1990savitzky}, PyWavelets method~\cite{lee2019pywavelets}, Wiener filter~\cite{chen2006new}, and Total-Variation denoising~\cite{chambolle2004algorithm}, respectively. The shaded areas indicate the standard deviation overlaid with the scatter plot. (b) MAPE of denoised signals using five methods versus MAPE of noisy signals. The shaded areas indicate the standard deviation overlaid with the scatter plot.}
  \label{fig:ensemble_result_comparison}
\end{figure}

To ensure a reliable evaluation of our denoising models, we employed a partitioning scheme that split the dataset into three subsets: training, holdout validation, and testing, in a 60:20:20 ratio. All neural network architectures were initialized with random weights and trained from scratch using a mini-batch size of 256 and a maximum of 500 iterations. Hyperparameters, including $\lambda_{\textit{LSTM}}$, $\lambda_{\textit{CNN}}$, $\lambda_{\textit{NN}}$ and $\lambda$, were selected based on their performance on the validation set. Our proposed denoising method was evaluated and compared to other algorithms using metrics such as Peak Signal-to-Noise Ratio (PSNR) and Weighted Mean Absolute Percentage Error (WMAPE). WMAPE is computed using the noise-free signal value as weights, which helps to alleviate the issue of error amplification for signals with small magnitudes. Fig.~\ref{fig:ensemble_result} (a) presents noisy signal and denoised signal versus a noise-free signal, where the dispersion illustrates the error distributions. The plot demonstrates the effectiveness of our method for reducing errors in the denoised signal, particularly for two selected ranges of noise level. The inset of Figure~\ref{fig:ensemble_result} (a) shows that the error for the noisy image increases significantly as the noise level increases. In contrast, the denoised signal generated by our model exhibits consistently low and stable error distributions. However, we observed mild heteroskedasticity, where smaller amplitudes (-0.2 to 0.2) were associated with larger observed errors~\cite{white1980heteroskedasticity}. To evaluate the robustness of our proposed denoising method, we conducted experiments by varying the noise level $\sigma_{\epsilon}$ from 0 to 0.2 and analyzing the trends of PSNR and WMAPE. Figure~\ref{fig:ensemble_result} (b) demonstrates that our model maintains consistent denoising performance across a wide range of noise levels, with PSNR values above 34 dB and WMAPE below 12\% even for $\sigma_{\epsilon}$ = 0.2. However, as the noise level increased, we observed an increase in error variance, which can compromise the overall performance of the model. 

To compare the performance of our proposed denoising method with existing signal denoising models, including the Savitzky-Golay filter~\cite{press1990savitzky}, PyWavelets method~\cite{lee2019pywavelets}, Wiener filter~\cite{chen2006new}, and Total-Variation denoising~\cite{chambolle2004algorithm}, we used PSNR, SNR and MAPE as key indicators. As illustrated in Figure~\ref{fig:ensemble_result_comparison}, our proposed model (designated as "Ensemble" in the figure) exhibits superior performance compared to the other models across the entire range of examined noise levels, as evidenced by its higher PSNR and lower WMAPE. However, for signals with very high levels of noise, the variance of the error may compromise our model's performance. Notably, the Total-Variance method was found to be closest to our model for signals with low levels of noise. The PyWavelets method demonstrates a consistent performance, manifested by the low variance of PSNR and MAPE for each noise level; however, the effectiveness of denoising may fall outside acceptable tolerances, which may be improved through further fine-tuning of the selection of wavelets and parameters~\cite{ngui2013wavelet}. In Table~\ref{table:comparison_models}, we compare the performance of different denoising algorithms on a set of testing datasets of sythetic structural vibration dataset holistically, using two different levels of noise: $\sigma_{\epsilon}$ = 0.1 and $\sigma_{\epsilon}$ = 0.2. The performance of each algorithm is evaluated based on three metrics: PSNR, SNR, and WMAPE. Among the algorithms compared in the table, our proposed method (labeled as "Ensemble" in the table) outperforms the others in terms of all the applied metrics, achieving an average PSNR of 38.0 dB for the first noise level and 35.8 dB for the second noise level. Our method also achieves competitive performance in terms of SNR and WMAPE, with an average SNR of 25.8 dB and 7.6 WMAPE for the first noise level, and 23.6 dB and 9.7 WMAPE for the second noise level, respectively. Compared to the other denoising algorithms in the table, Savgol achieves the lowest PSNR and SNR, while PyWavelets and Wiener perform similarly in terms of PSNR and SNR but have higher WMAPE values. Total-Variance method achieves competitive performance in terms of PSNR and WMAPE, but has a lower SNR than our proposed method. Our proposed denoising method exhibits exceptional performance compared to other algorithms on the tested datasets, owing to its specialized training on signals governed by structural dynamics, as well as the stacking ensemble architecture leveraging the frequency and time domain features; in contrast, other algorithms are designed for generic signal denoising scenarios.

\begin{table}[h]
\centering
\caption{Comparison of Models}
\label{table:comparison_models}
\begin{tabular}{lcccccc}
    \toprule
Model & \multicolumn{3}{c}{$\sigma_{\epsilon}$ = 0.1} & \multicolumn{3}{c}{$\sigma_{\epsilon}$ = 0.2} \\
 & PSNR & SNR & WMAPE & PSNR & SNR & WMAPE \\
 & [dB] & [dB] & [\%] & [dB] & [dB] & [\%] \\
    \toprule
Ensemble & 38.0 & 25.8 & 7.6 & 35.8 & 23.6 & 9.7 \\
Total-Variance & 31.2 & 19.0 & 15.8 & 24.2 & 11.8 & 29.4 \\
Wiener & 26.3 & 13.9 & 28.2 & 20.1 & 7.7 & 46.2 \\
Savgol & 27.6 & 15.4 & 24.1 & 22.2 & 10.1 & 40.4 \\
PYWT & 24.4 & 13.4 & 25.7 & 21.5 & 10.4 & 37.8 \\
    \toprule
\end{tabular}
\end{table}

\section{Conclusion}

Our research proposes a hybrid CNN-RNN tracking ensemble model for the denoising of single and multiple structural vibration signals, commonly encountered in civil, mechanical and bioengineering. The model is composed of three stages: a preprocessing stage, a CNN-RNN hybrid modeling stage, and an ensemble stage. The preprocessing stage extracts features from the input signal, such as frequency and time domain characteristics through FFT and wavelet transform. The CNN-RNN hybrid modeling, consisting of a bi-directional LSTM and CNN leverages these features to learn the underlying patterns and condensed feature representation. Finally, the ensemble stage aggregates the outputs of multiple models and time series to further improve performance. We tested the proposed model on several test datasets with varying levels of noise. The model outperformed existing algorithms across all datasets, as demonstrated by its superior performance in terms of PSNR, SNR, and WMAPE. Our model's success can be attributed to its specialized training on signals regulated by structural dynamics, which allowed it to learn more relevant features than previous models. Additionally, the innovative modeling architecture of the CNN-RNN hybrid stage contributed to the model's improved performance. Future work may involve the application of the proposed model to other types of vibration signals in bioengineering, as well as exploring the potential of the model in other fields such as structural health monitoring and fault diagnosis.

\bibliographystyle{unsrt}
\bibliography{reference}

\end{document}